\documentclass[%
 reprint,
 amsmath,amssymb,
 aps,
]{revtex4-1}

\usepackage{graphicx}
\usepackage{dcolumn}
\usepackage{bm}

\usepackage{subfigure}

\begin{document}

\preprint{APS/123-QED}

\preprint{APS/123-QED}

\title{Binary full adder, made of fusion gates,  in sub-excitable Belousov-Zhabotinsky system}

\author{Andrew Adamatzky}%
 \email{andrew.adamatzky@uwe.ac.uk}
\affiliation{%
 Unconventional Computing Centre, University of the West of England, Bristol, UK\\
}%

\date{\today}

\begin{abstract}
\noindent
In an excitable thin-layer Belousov-Zhabotinsky (BZ) medium a localised perturbation leads to formation of  omnidirectional target or spiral waves of excitation. A sub-excitable BZ medium responds to asymmetric local perturbation by producing travelling localised excitation wave-fragments, distant relatives of dissipative solitons.  The size and life span of an excitation wave-fragment depend on the illumination level of the medium. Under the right conditions the wave-fragments conserve their shape and velocity vectors for extended time periods.  We interpret the wave-fragments as values of Boolean variables. When two or more wave-fragments collide they annihilate or merge into a new wave-fragment.  States of the logic variables, represented by the wave-fragments, are changed in the result of the collision between the wave-fragments. Thus, a logical gate is implemented. Several theoretical designs and experimental laboratory implementations of Boolean logic gates have been proposed in the past but little has been done cascading the gates into binary arithmetical circuits. We propose a unique design of a binary one-bit full adder based on a fusion gate.  A fusion gate is a two-input three-output logical device which calculates conjunction of the input variables and conjunction of one input variable with negation of another input variable. The gate is made of three channels: two channels cross each other at an angle, third channel starts at the junction.  The channels contain BZ medium. When two excitation wave-fragments, travelling towards each other along input channels, collide at the junction they merge into a single wave-front travelling along the third channel. If there is just one wave-front in the input channel, the front continues its propagation undisturbed. We make a one-bit full adder by cascading two fusion gates. We show how to cascade the adder blocks into a many-bit full adder. We evaluate feasibility of our designs by simulating evolution of excitation in the gates and adders using numerical integration of Oregonator equations. 
\end{abstract}

\maketitle

\section{Introduction}

Information processing potential of a spatially extended, thin layer, Belousov-Zhabotinsky (BZ) medium~\cite{belousov1959periodic, zhabotinsky1964periodic} was firstly demonstrated in late 1980s in pioneer works by Kuhnert, Agladze and Krinsky on experimental laboratory implementation of spatial memory devices and image processing prototypes in the excitable chemical medium~\cite{kuhnert1986new, kuhnert1989image}. Their findings, albeit with a ten years delay, ignited a series of well founded theoretical and experimental works on realisation of computing devices in BZ medium. These include logical gates implemented in geometrically constrained BZ 
medium~\cite{steinbock1996chemical, sielewiesiuk2001logical}, approximation of shortest path by excitation waves~\cite{steinbock1995navigating, rambidi2001chemical, adamatzky2002collision}, memory in BZ micro-emulsion \cite{kaminaga2006reaction}, information coding with frequency of oscillations~\cite{gorecki2014information}, onboard controllers for robots~\cite{adamatzky2004experimental, yokoi2004excitable, DBLP:journals/ijuc/Vazquez-OteroFDD14}, chemical diodes~\cite{DBLP:journals/ijuc/IgarashiG11}, BZ neuromorphic architectures~\cite{gorecki2009information, gentili2012belousov, stovold2012simulating, takigawa2011dendritic}, wave-based counters~\cite{gorecki2003chemical}, and other aspects of information processing in excitable chemical systems~\cite{DBLP:journals/ijuc/YoshikawaMIYIGG09, escuela2014symbol, gruenert2014understanding, gorecki2015chemical}.

So far no complete arithmetical circuit has ever been implemented in BZ in experimental conditions. Some preparatory steps have  been done. They include simulation and experimental laboratory realisation of basic logical gates~\cite{adamatzky2004collision, adamatzky2007binary, toth2010simple, adamatzky2011towards} and generators of mobile localizations (they can play a role of constant {\sc True}) in light-sensitive BZ medium~\cite{de2009implementation} and adaptive design of simple logical gates using machine learning techniques~\cite{toth2009experimental}. The main reason for such slow experimental progress is instability of localised excitation wave-fragments: the waves are  either collapse or expand unless some external periodical stimulation is applied.  First one-bit half-adder in geometrically-constrained light-sensitive BZ medium is proposed in~\cite{adamatzky2010slime}, however the implementation required dynamical update of the illumination level to prevent excitation wave-fragments from collapsing or expanding. Models of multi-bit binary adder, decoder and comparator in BZ are proposed  
in~\cite{sun2013multi, zhang2012towards, suncrossover, digitalcomparator}. These architectures employ such crossover structures as T-shaped coincidence detectors~\cite{gorecka2003t} and chemical diodes~\cite{DBLP:journals/ijuc/IgarashiG11} that heavily rely on heterogeneity of geometrically constrained space. 

 Indeed by controlling excitability~\cite{igarashi2006chemical}  in different loci of the medium we can achieve impressive results, as it is demonstrated in works related to polymorphic logical gates~\cite{adamatzky2011polymorphic}, analogs of dendritic trees~\cite{takigawa2011dendritic}  and, particularly, implementation of four-bit input, two-bit output integer square root circuits based on alternating `conductivity' of junctions between channels~\cite{stevens2012time}, however experimental prototypes show a high degree of instability and sensitivity to environmental conditions. 
In present paper we try to overcome the problem of excitation wave-fragments instability via combining geometrical-constraining and collision-based approaches.

The paper  is structured as follows. We define Oregonator model of BZ medium and discuss parameters of simulation in Sect.~\ref{model}. In Sect.~\ref{twowaves} we outline differences in wave-front behaviour in excitable and sub-excitable BZ medium. The fusion gate is designed and its behaviour is simulated and analysed in Sect.~\ref{fusiongatesection}. Section~\ref{halfaddersection} presents architecture of a one-bit half adder made of a fusion gate. Two fusion gates are cascaded into a one-bit full adder in Sect.~\ref{fulladdersection}. In Sect.~\ref{cascading} we show how to cascade one-bit full adders into a many-bit full adder. Feasibility of the approach is discussed in Sect.~\ref{discussion}.

\section{Model}
\label{model}

We use two-variable Oregonator equation~\cite{field1974oscillations} adapted to a light-sensitive 
Belousov-Zhabotinsky (BZ) reaction with applied illumination~\cite{beato2003pulse}:
\begin{eqnarray}
  \frac{\partial u}{\partial t} & = & \frac{1}{\epsilon} (u - u^2 - (f v + \phi)\frac{u-q}{u+q}) + D_u \nabla^2 u \nonumber \\
  \frac{\partial v}{\partial t} & = & u - v 
\label{equ:oregonator}
\end{eqnarray}
The variables $u$ and $v$ represent local concentrations of an activator, or an excitatory component of BZ system, and an inhibitor, or a refractory component. Parameter $\epsilon$ sets up a ratio of time
scale of variables $u$ and $v$, $q$ is a scaling parameter depending
on rates of activation/propagation and inhibition, $f$ is a stoichiometric coefficient. 
Constant $\phi$ is a rate of inhibitor production.  In a light-sensitive BZ $\phi$ represents the rate of inhibitor production proportional to intensity of illumination (\ref{equ:oregonator}).

We integrate the system using Euler method with five-node Laplace operator, time step $\Delta t=0.001$ and grid point spacing $\Delta x = 0.25$, $\epsilon=0.02$ (see Sect.~\ref{twowaves}), $f=1.4$, $q=0.002$.   The parameter $\phi$ characterises excitability of the simulated medium. 

To generate excitation waves of wave-fragments we perturb the medium by square solid domains of excitation, $10 \times 10$ sites in state $u=1.0$; if different shape of perturbation was used we indicate this in the captions of the figures.

The medium is excitable and exhibits `classical' target waves when $\phi=0.05$ and the medium is sub-excitable with propagating localizations, or wave-fragments, when $\phi=0.0766$. Time lapse snapshots provided in the paper were recorded at every 150 time steps, we display sites with $u >0.04$.

The model has been repeatedly verified by us in experimental laboratory studies of BZ system, and  the perfect match between the model and the experiments was demonstrated in \cite{adamatzky2007binary, de2009implementation, toth2010simple, adamatzky2011towards}.

We adopt the following symbolic notations. Boolean variables $x$ and $y$ take values `0', logical {\sc False}, and `1', logical {\sc True}; $xy$ is a conjunction (operation {\sc and}),
 $x\oplus y$ is exclusive disjunction (operation {\sc xor}), $\overline{x}$ is a negation of variable $x$ ({\sc not}). Disjunction ({\sc OR}) of variables $x$ and $y$ is $x+y$.

\section{Excitable versus sub-excitable media}
\label{twowaves}

\begin{figure}[!tbp]
\subfigure[]{\includegraphics[width=0.4\linewidth]{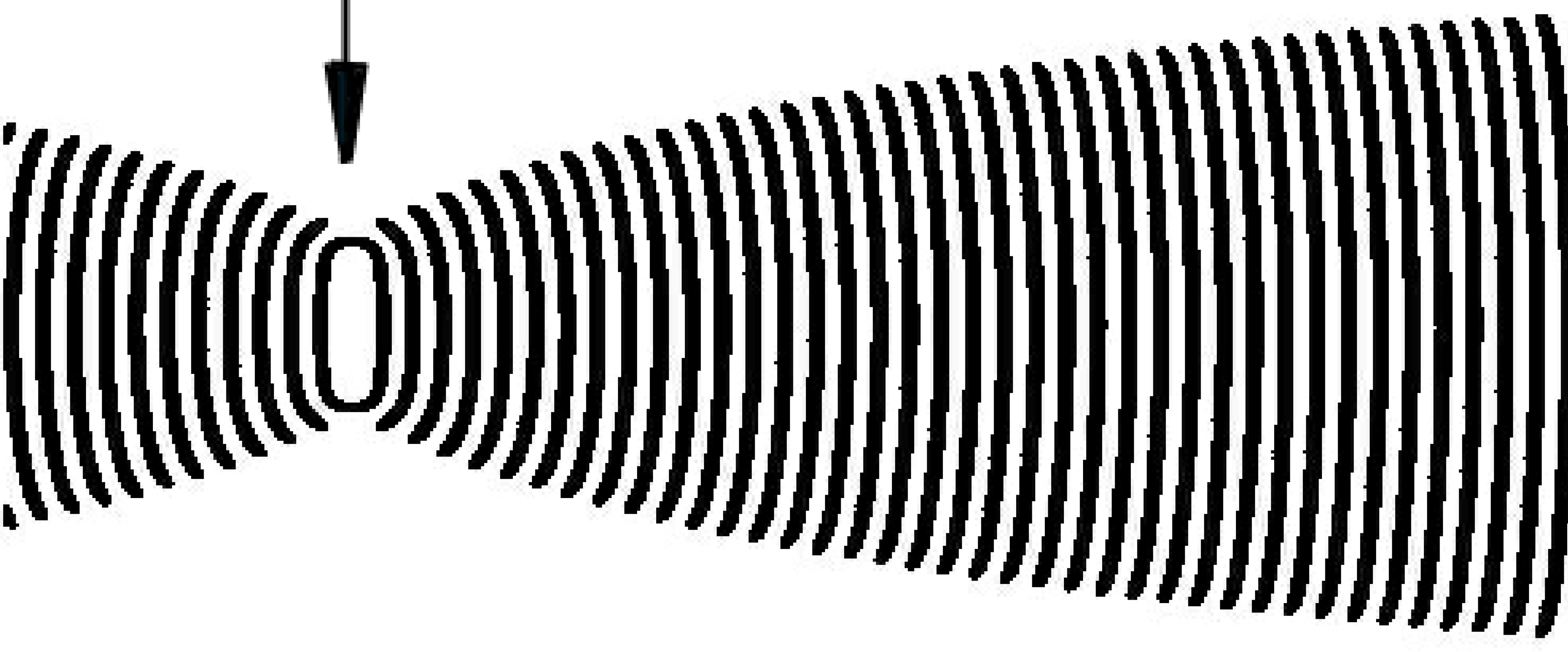}}\\
\subfigure[]{\includegraphics[width=0.7\linewidth]{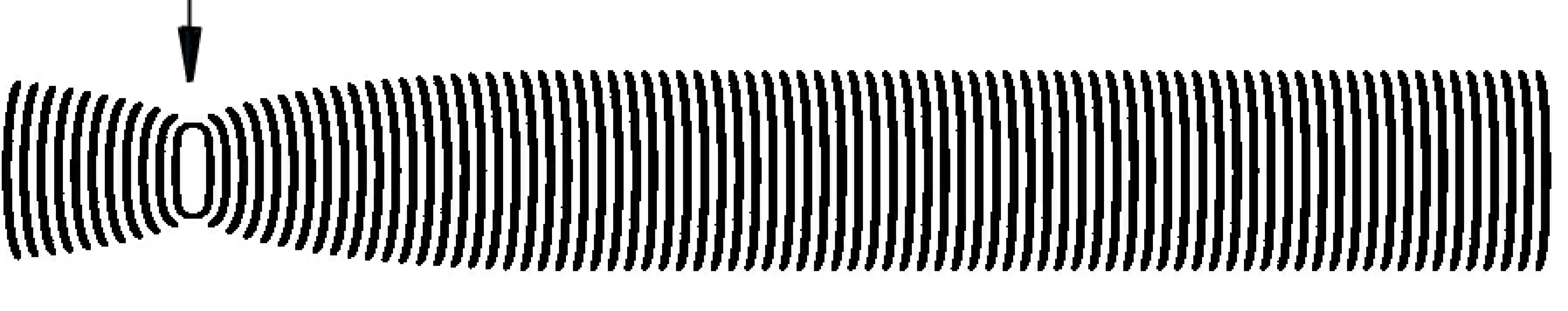}}
\subfigure[]{\includegraphics[width=0.7\linewidth]{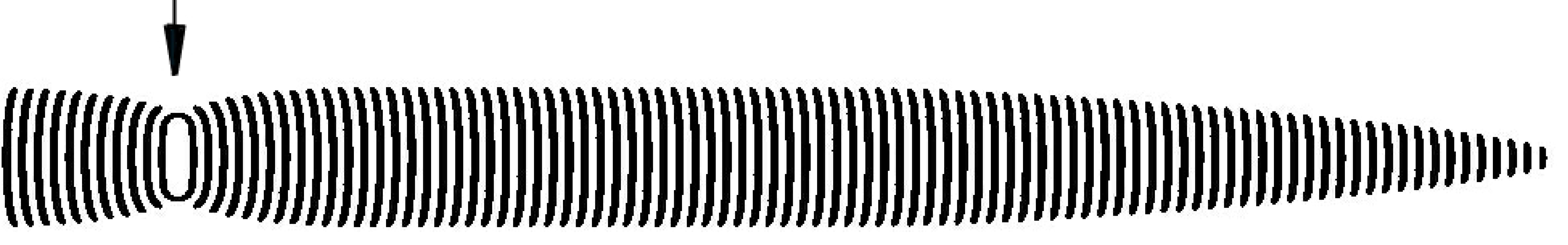}}
\caption{Time lapsed snapshots of wave-fragments propagating in
  simulated BZ medium in (a)~near lower threshold of excitability, $\phi=0.079$,
  (b)~sub-excitable mode, $\phi=0.07905$, (c)~non-excitable mode,
  $\phi=0.0791$. The media were perturbed by rectangular north-south
  elongated domains of excitation, $3 \times 40$ sites in state
  $u=1.0$.  Sites of initial segment-wise perturbation are shown by
  arrows. Grid size is 1125$\times$250 nodes. 
  The pictures are not snapshots of many wave fronts generated at the initial stimulation point. These are time lapsed snapshots of two waves (one propagating right, another left) recorded every $150^{th}$ step of
  numerical integration.}
\label{fragments}
\end{figure}

Given initial asymmetric excitation domain a wave-fragment is formed. The fragment's velocity vector is a normal to longest side of the perturbation domain. The medium is excited by rectangular domains of perturbed sites. The perturbation domains are elongated along north-south axis therefore wave-fragments generated propagate west and east. In excitable medium, with low threshold of excitation, every resting site neighbouring excited site becomes excited. Therefore literally a point-wise excitation is enough to generate full-scale omni-directional wave. In sub-excitable medium a resting site needs several excited neighbours to get excited. Initially asymmetric wave will cause excitation of resting sites residing nearby middle of the wave-front only. Therefore the excitation wave-fragment `grows' from its centre and `dies' at its ends. This is why wave-fragments in Fig.~\ref{fragments} propagate only east and west and shrink or just slightly expand along south-north direction.

Depending on medium's excitability wave-fragments may expand (Fig.~\ref{fragments}a), keep their shape for a long time (Fig.~\ref{fragments}b) or collapse (Fig.~\ref{fragments}c).
In ideal situation, assuming that wave-fragments keep their shapes indefinitely, we can implement a collision-based computing circuit of any depth subject to space availability.  However in reality,
particularly in conditions of chemical laboratory experiments, wave-fragments are very unstable. It is almost impossible to keep a medium at the precise level of sub-excitability (Fig.~\ref{fragments}b), and almost any wave-fragment will expand (Fig.~\ref{fragments}a) or collapse (Fig.~\ref{fragments}c).  

It is possible to keep wave-fragment from collapsing or exploding by periodically changing excitability of the medium~\cite{sakurai2002design}.  When a wave-fragment expands we increase illumination level thus decreasing the medium's excitability. When the wave-fragment starts to collapse the illumination is decreased, the medium's excitability increases and the wave-fragment expands.  In designs of logical circuits when channels are represented by shadowed areas and other parts of medium with illuminated areas, it would be difficult to keep periodicity of illumination and synchronise signals are the same time. Therefore, we adopted the following approach. We keep the medium in excitable mode inside channels and in sub-excitable mode at the junctions. Thus we do not worry about collapsing of wave-fronts inside channels, yet we keep wave-fronts localised when they cross junctions between channels.

\section{Fusion gate}
\label{fusiongatesection}

\begin{figure}[!tbp]
\includegraphics[width=0.7\linewidth]{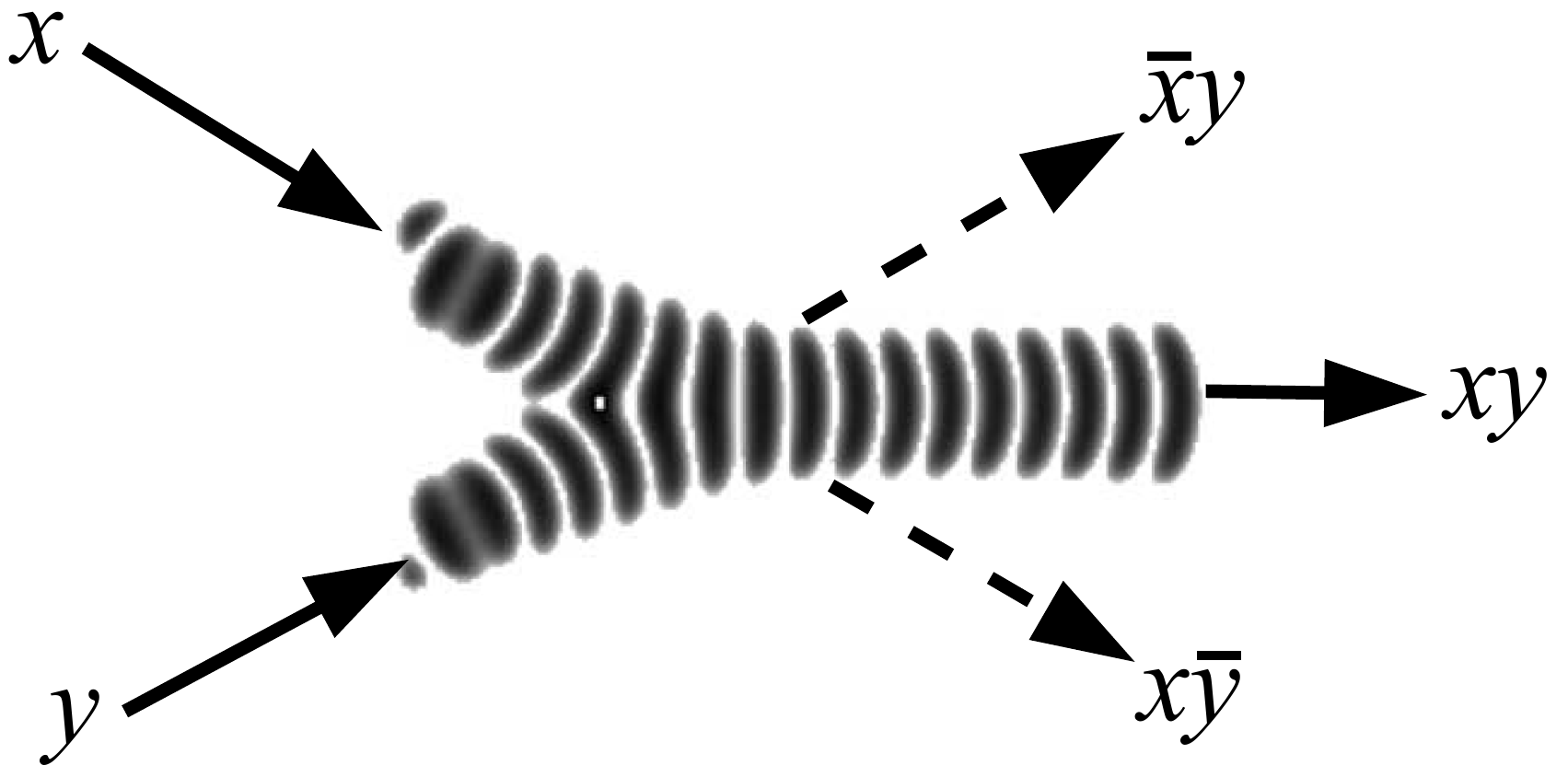}
\caption{Time lapsed overlays of the fusion of two excitation wave-fragments. One fragment is travelling from north-west to south-east, another fragment from south-west to north-east. The wave-fragments collide and fuse into a new localised excitation travelling east. }
\label{collision}
\end{figure}

When two wave-fragments, of equal size,  travelling towards each other in a sub-excitable medium collide head-on, 
they annihilate.  When the wave-fragments collide at an acute angle they fuse into a single wave-fragment (Fig.~\ref{collision}). If one of the wave-fragments was present another fragment would move along its original trajectory.

\begin{figure}[!tbp]
\subfigure[]{\includegraphics[width=0.3\linewidth]{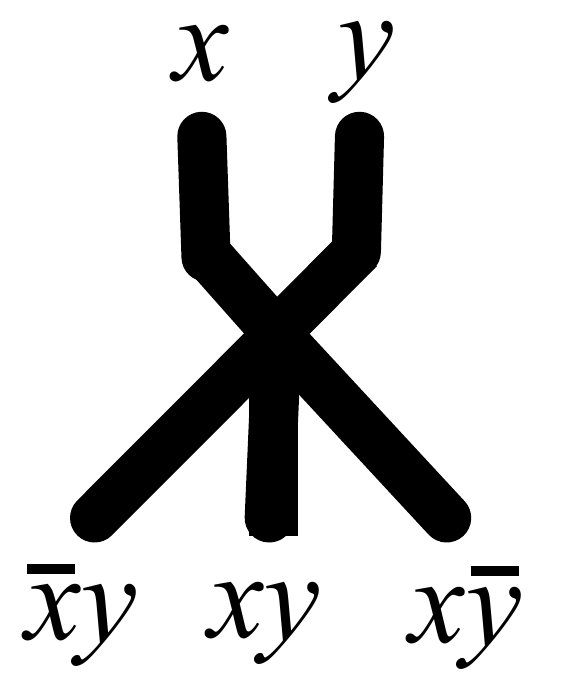}}\\
\subfigure[]{\includegraphics[width=0.44\linewidth]{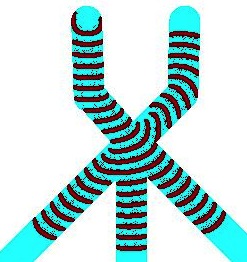}}
\subfigure[]{\includegraphics[width=0.49\linewidth]{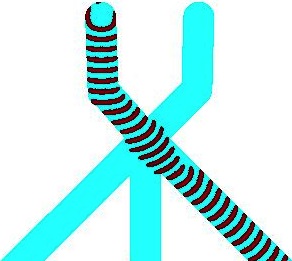}}
\subfigure[]{\includegraphics[width=0.39\linewidth]{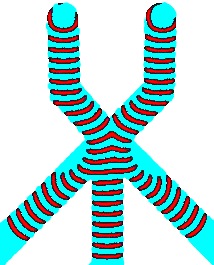}}
\subfigure[]{\includegraphics[width=0.49\linewidth]{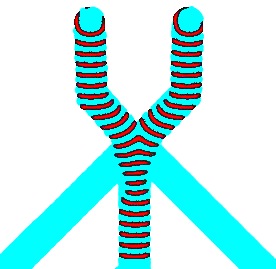}}
\caption{Fusion gate. 
(a)~Scheme: inputs are $x$ and $y$, outputs are $xy$, $\overline{x}y$, $x\overline{y}$.  
(a--d)~Time lapsed overlays of excitation waves.
(b)~Excitable mode, $x=1$, $y=0$.
(c~)~Sub-excitable mode, $x=1$, $y=0$.
(d~)~Excitable mode,  $x=1$, $y=1$.
(e)~Sub-excitable mode,  $x=1$, $y=1$.
}
\label{fusiongate}
\end{figure}

\begin{figure*}[!tbp]
\subfigure[$t=0.475$]{\includegraphics[width=0.25\linewidth]{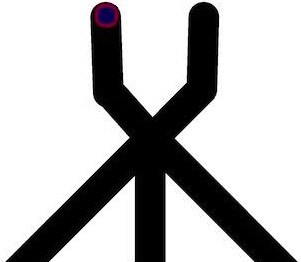}}
\subfigure[$t=6.91$]{\includegraphics[width=0.25\linewidth]{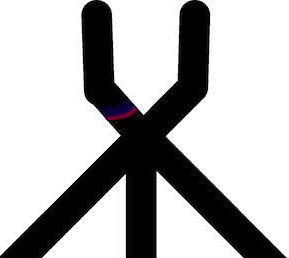}}
\subfigure[$t=8.76$]{\includegraphics[width=0.25\linewidth]{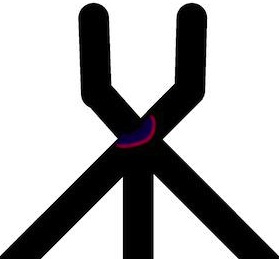}}
\subfigure[$t=9.955$]{\includegraphics[width=0.25\linewidth]{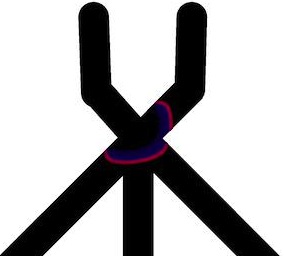}}
\subfigure[$t=11.21$]{\includegraphics[width=0.25\linewidth]{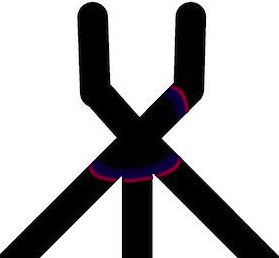}}
\subfigure[$t=13.785$]{\includegraphics[width=0.25\linewidth]{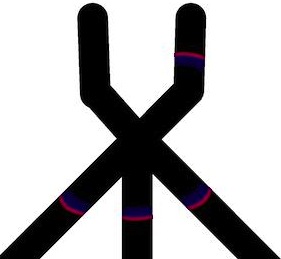}}\\
\subfigure[$t=0.37$]{\includegraphics[width=0.25\linewidth]{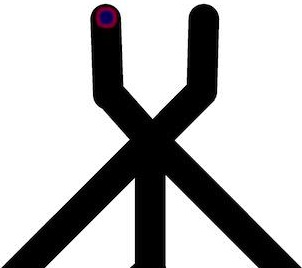}}
\subfigure[$t=7.14$]{\includegraphics[width=0.25\linewidth]{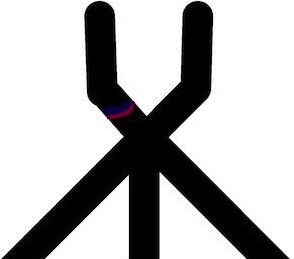}}
\subfigure[$t=8.785$]{\includegraphics[width=0.25\linewidth]{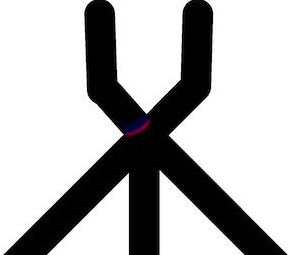}}
\subfigure[$t=10.135$]{\includegraphics[width=0.25\linewidth]{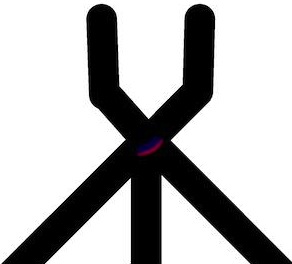}}
\subfigure[$t=11.31$]{\includegraphics[width=0.25\linewidth]{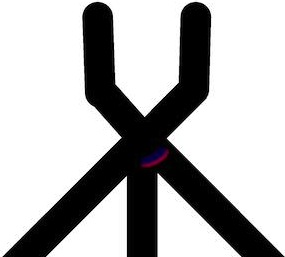}}
\subfigure[$t=13.605$]{\includegraphics[width=0.25\linewidth]{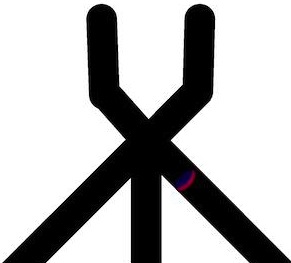}}
\caption{Snapshots of fusion gate for inputs $x=1$ and $y=0$. (a--f)~Excitable mode. (g--l)~Sub-excitable mode. Time is shown in captions. Excitable heads of the wave-fragments are shown in red. Refractory tail in blue. }
\label{fusiongatesnapshots}
\end{figure*}

This give us an idea of a fusion gate. The gate is built of channels containing BZ medium being in a sub-excitable mode. 
The gate has two input channels and three outputs channels (Fig.~\ref{fusiongate}a). When there is a wave-fragment present in the input channel $x$ ($y$) we assume that input variable $x$ ($y$) takes logical Boolean value {\sc True}, $x=1$ ($y=1$); otherwise Boolean value {\sc False}, $x=0$ ($y=0$).  

If the gate is in excitable mode  and initiated with input $x=1$ and $y=0$ (Fig.~\ref{fusiongate}b and Fig.~\ref{fusiongatesnapshots}ab), then wave-fragment propagating along input channel $x$ expands at the junction (Fig.~\ref{fusiongatesnapshots}cd). Excitation wave-front then enters all other channels (Fig.~\ref{fusiongate}b  and Fig.~\ref{fusiongatesnapshots}ef). Such gate implements only multiplication of a signal and it is not useful for a realisation of logical functions.  If the gate is in a sub-excitable mode and  initiated with $x=1$ and $y=0$ (Fig.~\ref{fusiongate}c and Fig.~\ref{fusiongatesnapshots}gh) then the wave-fragment does not expand at the junction (Fig.~\ref{fusiongatesnapshots}ij) but continues through the junction along its original trajectory  
Fig.~\ref{fusiongatesnapshots}kl).

\begin{figure*}[!tbp]
\subfigure[$t=0.475$]{\includegraphics[width=0.25\linewidth]{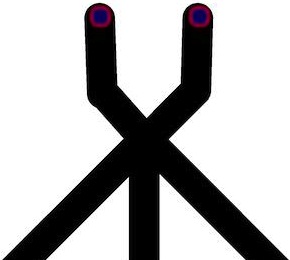}}
\subfigure[$t=0.475$]{\includegraphics[width=0.25\linewidth]{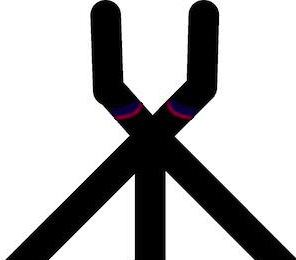}}
\subfigure[$t=0.475$]{\includegraphics[width=0.25\linewidth]{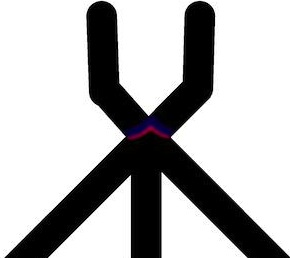}}
\subfigure[$t=0.475$]{\includegraphics[width=0.25\linewidth]{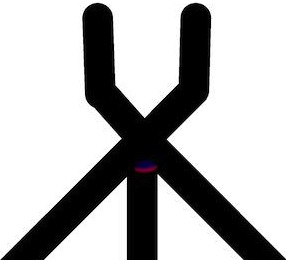}}
\subfigure[$t=0.475$]{\includegraphics[width=0.25\linewidth]{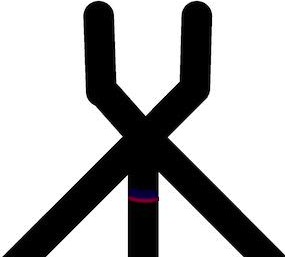}}
\caption{Snapshots of fusion gate for inputs $x=1$ and $y=1$. The system is in sub-excitable mode. Time is shown in captions. Excitable heads of the wave-fragments are shown in red. Refractory tail in blue. }
\label{fusiongatemerging}
\end{figure*}

If the gate was in excitable mode  and initiated with input $x=1$ and $y=1$ (Fig.~\ref{fusiongate}d) the wave-fragments collide, fuse and expand. The excitation wave-fronts enter all three output channels. In sub excitable mode the wave-fragments fuse   (Fig.~\ref{fusiongate}e) into a single localised wave-fragment which enters the central output channel (Fig.~\ref{fusiongatemerging}). 

Thus, assuming that presence of a wave-fragment in an output channel symbolises logical variable in state {\sc True} at the output channel, we conclude that the fusion gate, being in sub-excitable mode, implements two-input-three-output Boolean logic gate $\langle x, y \rangle \rightarrow \langle  \overline{x}y, xy, x\overline{y} \rangle$ (Fig.~\ref{fusiongate}a). The fusion gate is a key component of a one-bit half-adder presented in next section.

\section{Half-Adder}
\label{halfaddersection}

\begin{figure}[!tbp]
\includegraphics[width=0.4\linewidth]{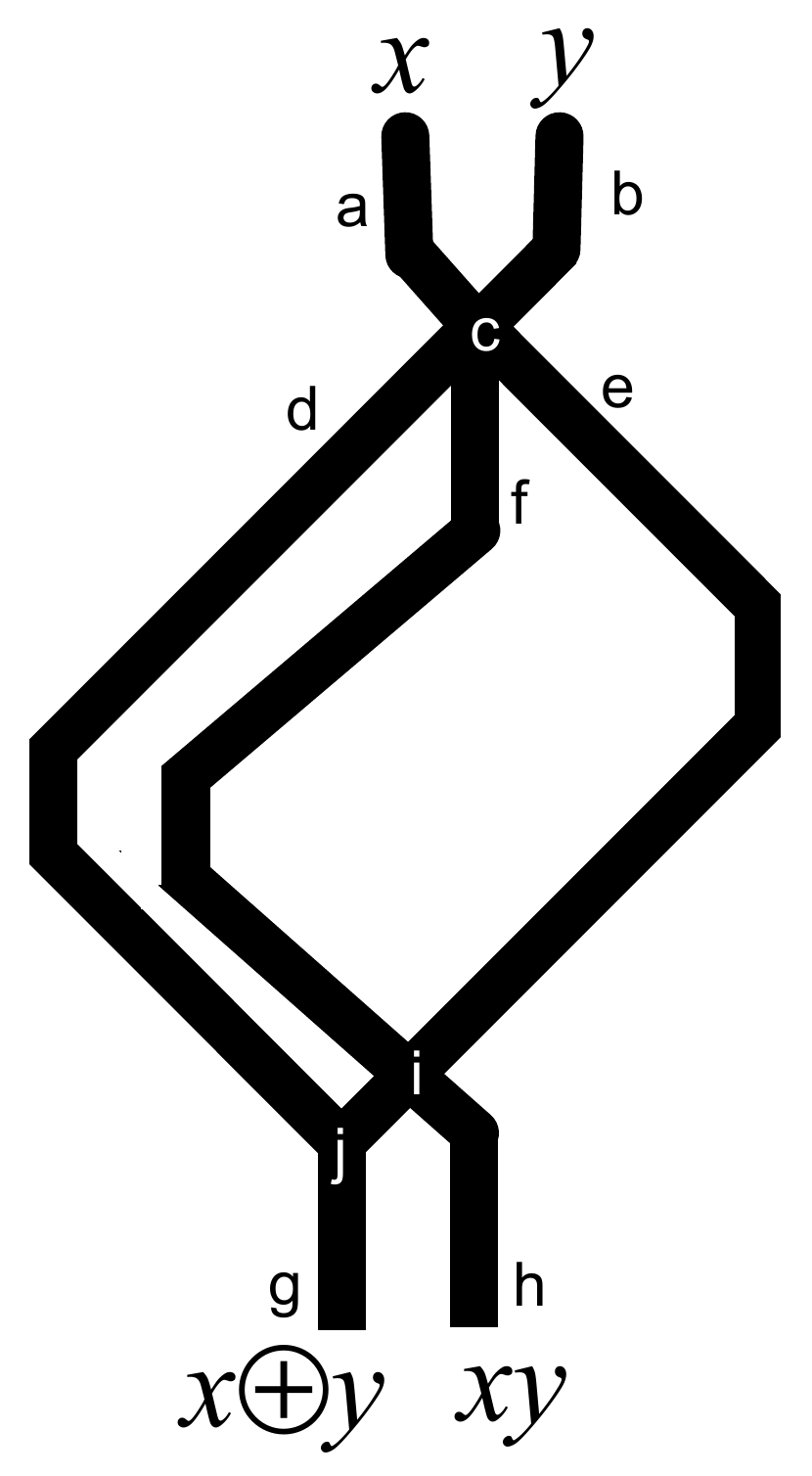}
\caption{A scheme of a one-bit half-adder: input channels are $a$ and $b$, output channels are $g$ and $h$, internal channels $d$, $e$, $f$, and junctions are $c$, $i$ and $j$. Input variables $x$ and $y$ are fed into channels $a$ and $b$, results $x \oplus y$ and $xy$ are read from channels $g$ and $h$.}
\label{halfadderscheme}
\end{figure}

A one-bit half-adder is a device with two inputs $x$ and $y$ and two outputs $xy$ (Carry out) and $x\oplus y$ (Sum).  The half-adder based on the fusion gate is shown in Fig.~\ref{halfadderscheme}. It consists of two input channels $a$ and $b$ and two output channels $g$ and $h$. Presence/absence of a wave-fragment in an input/output channel  symbolises logical {\sc True}/{\sc False} state of the input variable assigned to the channel. Synchronisation of signal wave-fragments is achieved geometrically: $|a|+|e|+|g| = |b|+|d|+|g|=|a|+|f|+|h|$, where $|\cdot |$ is a length of a channel (Fig.~\ref{halfadderscheme}).

\begin{figure}[!tbp]
\subfigure[]{\includegraphics[width=0.49\linewidth]{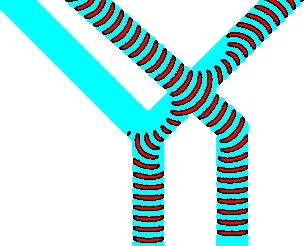}}
\subfigure[]{\includegraphics[width=0.45\linewidth]{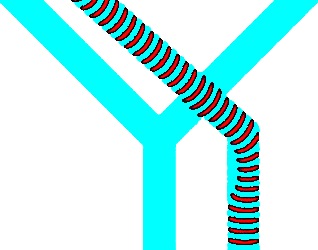}}
\subfigure[]{\includegraphics[width=0.49\linewidth]{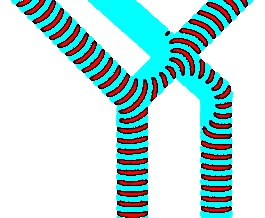}}
\subfigure[]{\includegraphics[width=0.45\linewidth]{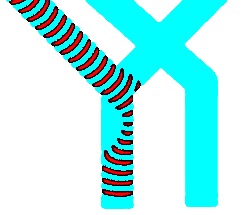}}
\caption{Time lapsed overlays of excitation waves propagating through junctions $i$ (ab) and $j$ (cd). 
(ab)~Wave-fragment approaches junction $i$ via channel $f$ (Fig.~\ref{halfadderscheme}). 
(cd)~Wave-fragment approaches junction $j$ via channel $d$ (Fig.~\ref{halfadderscheme}). 
(ac)~The system is in excitable mode. 
(bd)~The system is in sub-excitable mode.
}
\label{halfadderjunctions}
\end{figure}

The half-adder has three junctions $c$, $i$ and $j$ (Fig.~\ref{halfadderscheme}). The junction $c$ is key part of the  fusion gate (Fig.~\ref{fusiongate}a) and propagation of wave-fragments through this junction has been discussed in Sect.~\ref{fusiongate}. Propagation of waves via junctions $i$ and $j$ is illustrated in Fig.~\ref{halfadderjunctions} for excitable (Fig.~\ref{halfadderjunctions}ac) and sub-excitable (Fig.~\ref{halfadderjunctions}bd) modes. 

If the system was in excitable mode then wave-fragment entering junction $i$ via channel $f$ would expand and propagate into channels $e$ (above and below the junction) and $h$ (Fig.~\ref{halfadderjunctions}a). Similarly, the wave-fragment entering junction $j$ from channel $d$ expands and propagates into channels $j$, $e$ and $h$ (Fig.~\ref{halfadderjunctions}a). Signals are multiplied and thus, again, the device being in excitable mode is not useful for implementing logical functions. When the system is in sub-excitable mode a wave-fragment entering junction $i$ via channel $f$ stays localised and propagates only into channel $h$ (Fig.~\ref{halfadderjunctions}b). Similarly, a wave-fragment entering junction $j$ from channel $d$ propagates only into channel $g$ (Fig.~\ref{halfadderjunctions}d). There is no such combination of inputs which might lead to collision of two wave-fragments in junction $i$, therefore we do not consider such scenario. 

\begin{figure*}[!tbp]
\subfigure[]{\includegraphics[width=0.32\linewidth]{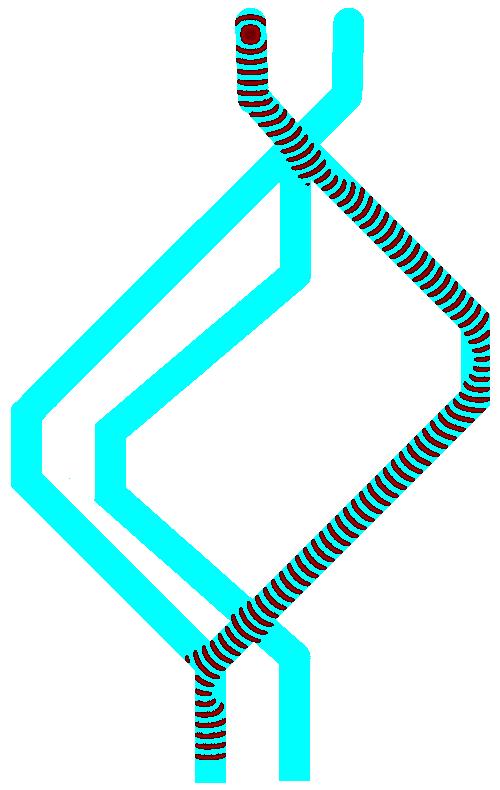}}
\subfigure[]{\includegraphics[width=0.32\linewidth]{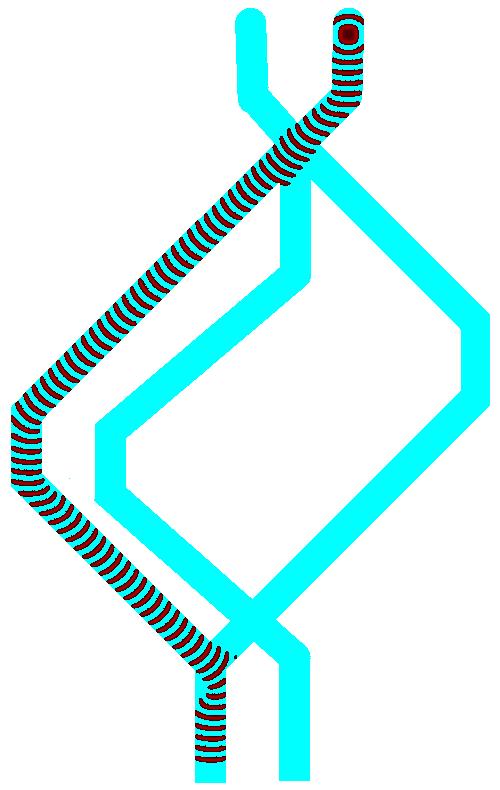}}
\subfigure[]{\includegraphics[width=0.32\linewidth]{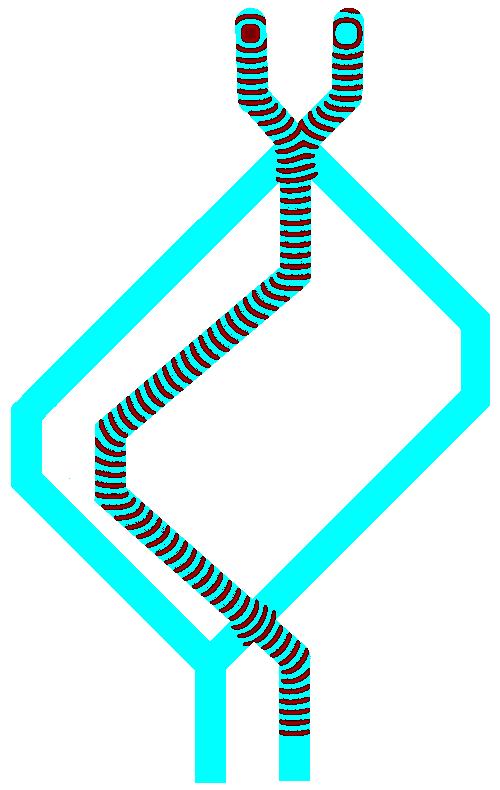}}
\caption{Time lapsed overlays of excitation waves propagation in the one-bit half-adder (Fig.~\ref{halfadderscheme}) for inputs (a)~$x=1$, $y=0$, (b)~$x=1$, $y=0$, (c~)~$x=1$, $y=1$. 
 Sites of initial segment-wise perturbation are visible as discs. 
 Grid size is 500$\times$790 nodes. The pictures are not snapshot of many wave fronts generated at the initial stimulation point. These are time lapsed snapshots of a single wave (ab) or two waves merging into a single wave (c) recorded every $150^{th}$ step of   numerical integration.
}
\label{halfadder}
\end{figure*}

The half-adder in action is shown in Fig.~\ref{halfadder}.  When inputs are $x=1$ and $y=0$ a wave-fragment is initiated in channel $a$. The wave stays localised (without expansion) when propagating via junction $c$ and enters channel $e$. The wave travels along the channel $e$ and through junction $i$ and $j$ and enters output channel $g$ (Fig.~\ref{halfadder}a). Similarly, when inputs are $x=0$ and $y=1$  a wave is initiated in channel $b$, propagates though junction $c$, along channel $d$, via junction $j$ and enters output channel $g$ (Fig.~\ref{halfadder}b). If both inputs takes value {\sc True} excitation wave-fragments are initiated in channels $a$ and $b$ simultaneously. The wave-fragments collide at junction $c$ and fuse into a single wave-fragment. This newly born wave-fragment propagates into channel $f$. The wave travels along channel $f$, passes though junction $i$ without expansion and enters the output channel $h$ (Fig.~\ref{halfadder}c). 

A wave-fragment appears in output channel $g$ only if either a wave-fragment was initiated either in input channel $a$ or input channel $b$. Therefore, the channel $g$ represents logical operation $x \oplus y$.  A wave-fragment appears in output channel $h$ only if wave-fragments were initiated in input channels $x$ and $y$. Therefore, the channel $h$ represents logical operation $xy$. Thus, we illustrated how the half-adder shown in Fig.~\ref{halfadderscheme} is derived.

\section{One-bit full adder}
\label{fulladdersection}

\begin{figure}[!tbp]
\includegraphics[width=0.45\linewidth]{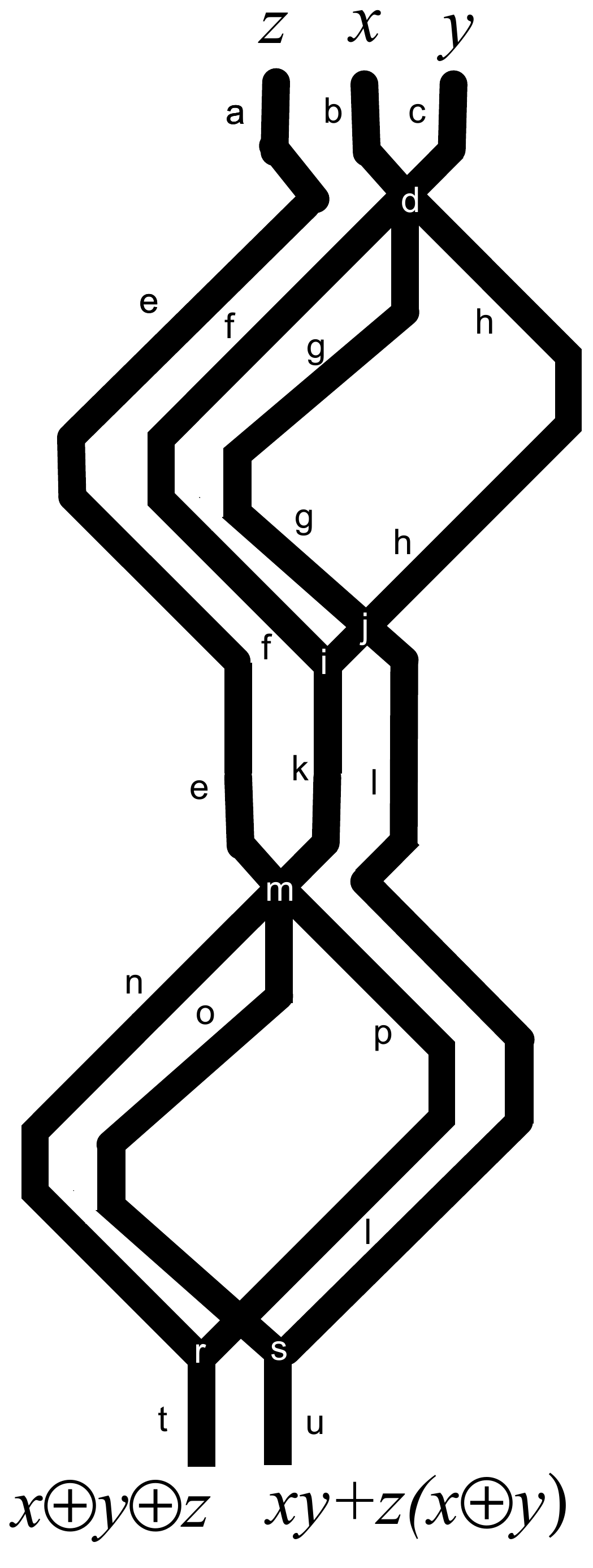}
\caption{A scheme of a one-bit full-adder: input channels are $a$, $b$ and $c$, output channels are $t$ and $u$, internal channels $e$, $f$, $g$, $h$, $k$, $l$, $n$, $o$, $p$, 
and junctions are $d$, $i$, $j$, $m$, $r$, $s$. Input variables $x$ and $y$ are fed in channels $b$ and $c$, Carry In is fed in channel $a$, results $x \oplus y \oplus z$ and $xy + z(x \oplus y)$ are read from channels $t$ and $u$.}
\label{fulladderscheme}
\end{figure}

A one-bit full adder is a device with three inputs $x$, $y$ and $z$ (Carry in) and two outputs  
$xy + z(x\oplus y)$ (carry out) and $x\oplus y\oplus z$ (Sum). The full adder is made of two half-adders  
(Fig.~\ref{fulladderscheme}) by cascading Carry in  into input channel of second half-adder, Sum of first half-adder into second input of second half-adder, and Carry out of first half-adder into {\sc or} junction with Carry out output channel of second half-adder. The one-bit full adder has three inputs channels $a$, $b$ and $c$ (they represent carry in $z$, and added bits $x$ and $y$) and two output channels $t$ and $u$ (they represent sum $x\oplus y\oplus z$ and carry out $xy + z(x\oplus y)$).  Synchronisation of signals is achieved geometrically: 
$|a|+|e|+|p|+|t| = |b|+|h|+|k|+|n|+|t|= |c|+|f|+|k|+|n|+|t|=|b|+|g|+|l|+|u|$. Six junctions $d$, $i$, $j$, $m$, $r$ and $s$ are types of junctions already discussed in Sect.~\ref{halfadder}.

\begin{figure*}[!tbp]
\subfigure[]{\includegraphics[width=0.3\linewidth]{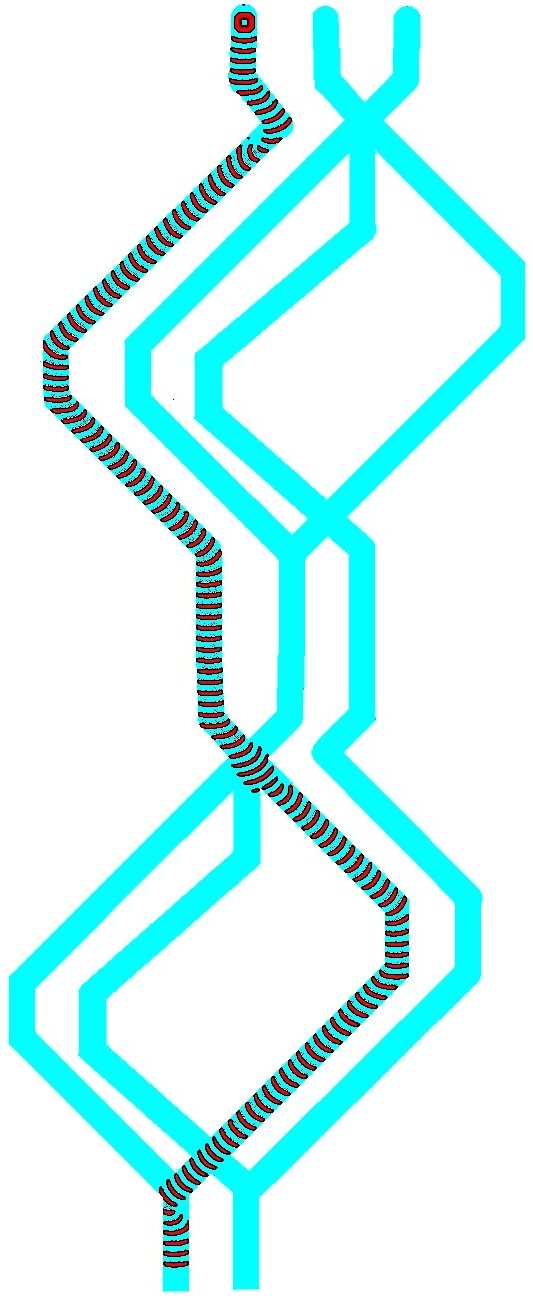}}
\subfigure[]{\includegraphics[width=0.3\linewidth]{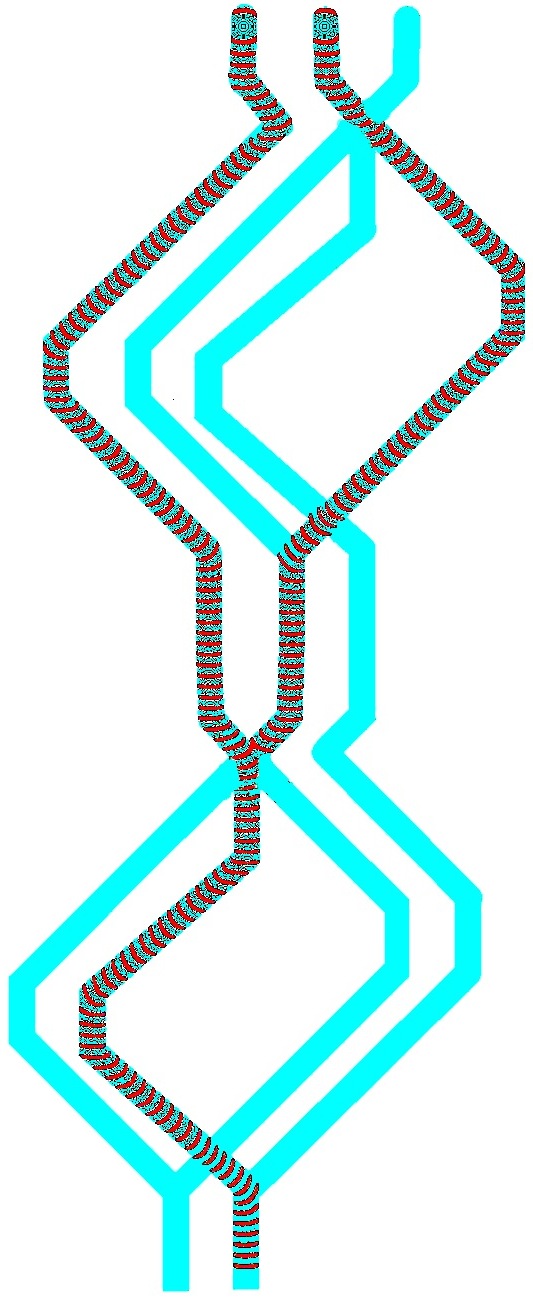}}
\subfigure[]{\includegraphics[width=0.3\linewidth]{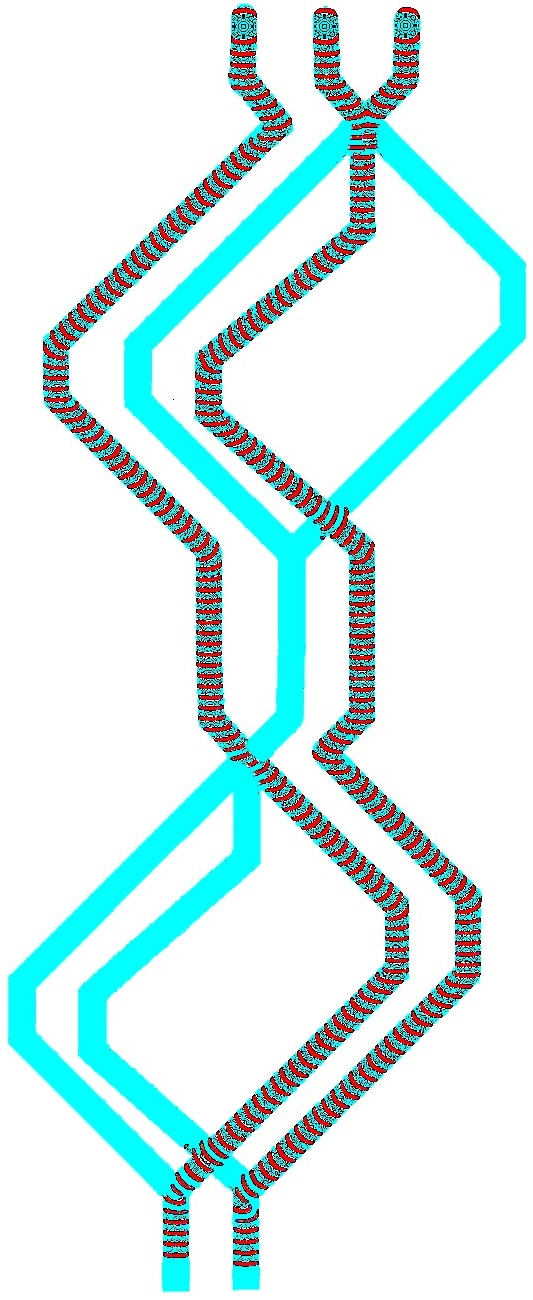}}
\caption{Time lapsed overlays of excitation waves propagation in the one-bit full adder (Fig.~\ref{fulladderscheme}) for inputs (a)~$x=0$, $y=0$, $z=1$, (b)~$x=1$, $y=0$, $z=1$, 
(c~)~$x=1$, $y=1$., $z=1$.  }
\label{addersnapshots}
\end{figure*}

\begin{table}[!tbp]
\caption{Schematic dynamic of one-bit full adder for all possible combinations of inputs. 
The adder, as shown in Fig.~\ref{fulladderscheme} is rotated to the left.}
\begin{tabular}{c|c|c||c|c|c}
\hspace{0.07cm} $x$ \hspace{0.07cm} & \hspace{0.07cm} $y$ \hspace{0.07cm}  & $C_{in}$ & $Sum$ & $C_{out}$ & Configuration \\ \hline
0 & 0 & 0 & 0 & 0 & \includegraphics[angle=90,scale=0.4]{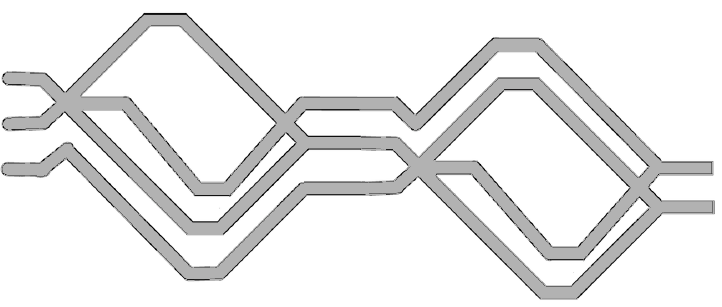}  \\ \hline
1 & 0 & 0 & 1 & 0 & \includegraphics[angle=90,scale=0.4]{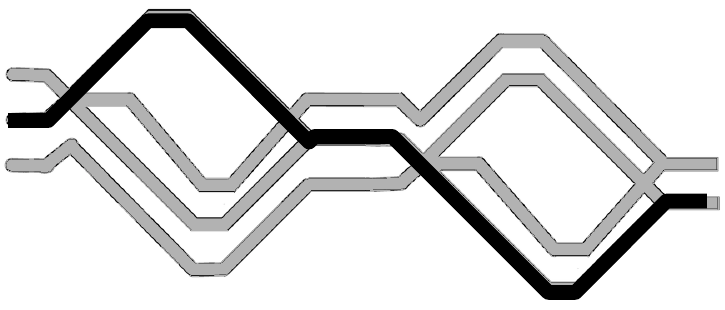}  \\ 
0 & 1 & 0 & 1 & 0 & \includegraphics[angle=90,scale=0.4]{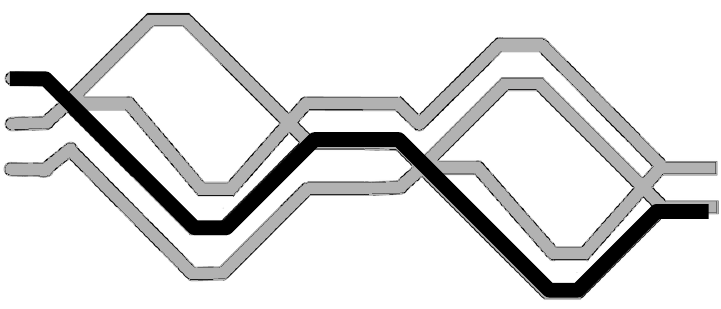}  \\ 
1 & 1 & 0 & 0 & 1 & \includegraphics[angle=90,scale=0.4]{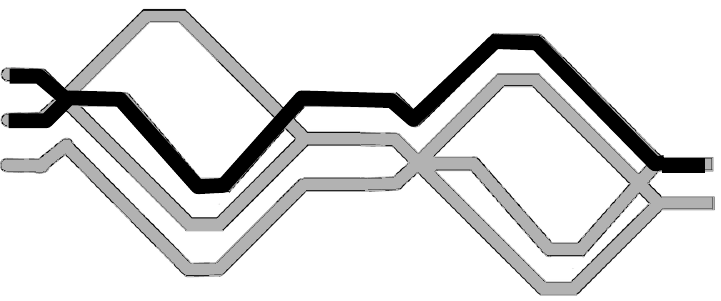}  \\
0 & 0 & 1 & 1 & 0 & \includegraphics[angle=90,scale=0.4]{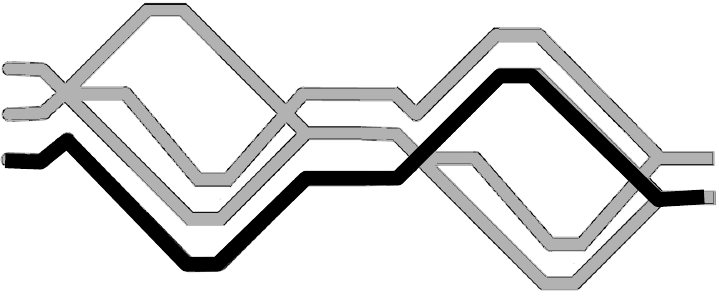}  \\ 
1 & 0 & 1 & 0 & 1 & \includegraphics[angle=90,scale=0.4]{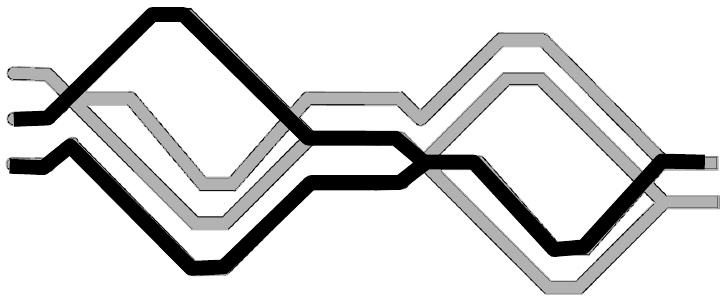}  \\ 
0 & 1 & 1 & 0 & 1 & \includegraphics[angle=90,scale=0.4]{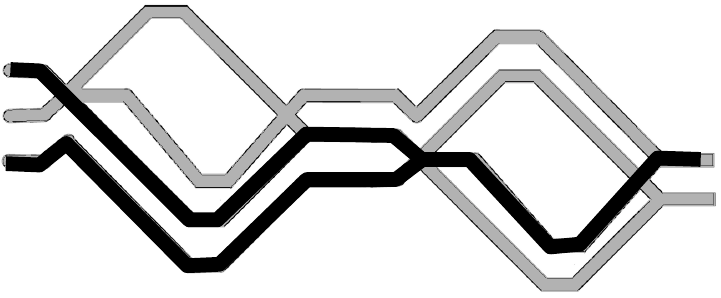}  \\ 
1 & 1 & 1 & 1 & 1 & \includegraphics[angle=90,scale=0.4]{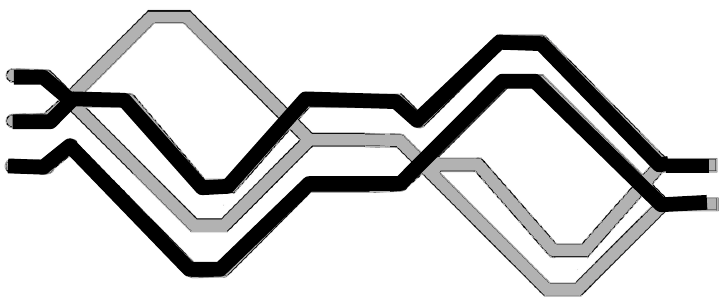}  \\ \hline
\end{tabular}
\label{tableadder}
\end{table}

The one-bit full adder in action is shown in Fig.~\ref{addersnapshots}.  When Carry in has bit up, $z=1$, and two inputs bits are down, $x=0$ and $y=z$ the wave-fragment is originated only in input channel $a$ (Fig.~\ref{addersnapshots}a). The wave travels along channel $e$, through junction $m$, enters channel $p$, propagates though junction $r$ and finishes its journey in output channel $t$.  For input tuple $z=0$, $x=1$ and $y=0$, an excitation wave-front is initiated in input channel $b$, propagates through junction $d$ into channel $h$, through junctions $j$ and $i$ into channel $k$, through junction $n$ into channel $n$, and finally enters the output channel $t$. When inputs are $z=0$,, $x=0$ and $y=1$ a wave-front initiated in input channel $c$ propagates through $d$, along $f$, through $i$, along $k$, through $m$, along $n$ and ends up in output channel $t$.    For combination of inputs $z=0$, $x=1$ and $y=1$ wave-fragments are initiated in input channels $b$ and $c$, they merge into a single wave-fragment at junction $d$. This newly born wave-fragment propagates along $g$, through $j$, along $l$, through $s$ and into output channel $u$. 
Dynamics of excitation waves for inputs $z=1$, $z=1$ and $y=0$ is shown in Fig.~\ref{addersnapshots}b. The wave-fragments merge into a single wave-fragment at junction $m$ and the newly born fragment propagates into output channel $u$.  When all three input has state bit up, $z=1$, $x=1$, $y=1$, the wave-fragment representing $z=1$ does not interact with the wave-fragment produced by merged wave-fragments representing $x=1$ and $y=1$  (Fig.~\ref{addersnapshots}c). Thus we have wave-fragments appearing on both output channels $t$ and $u$. 

Trajectories of wave-fragments for all possible combinations of inputs are shown in Tab.~\ref{tableadder}. The output channel $t$ symbolises Sum $x \oplus y \oplus z$ and channel $u$ Carry out $xy + z(x\oplus y)$. 

\section{Cascading one-bit full adder in many-bits full adder} 
\label{cascading}

\begin{figure}[!tbp]
\includegraphics[width=0.5\linewidth]{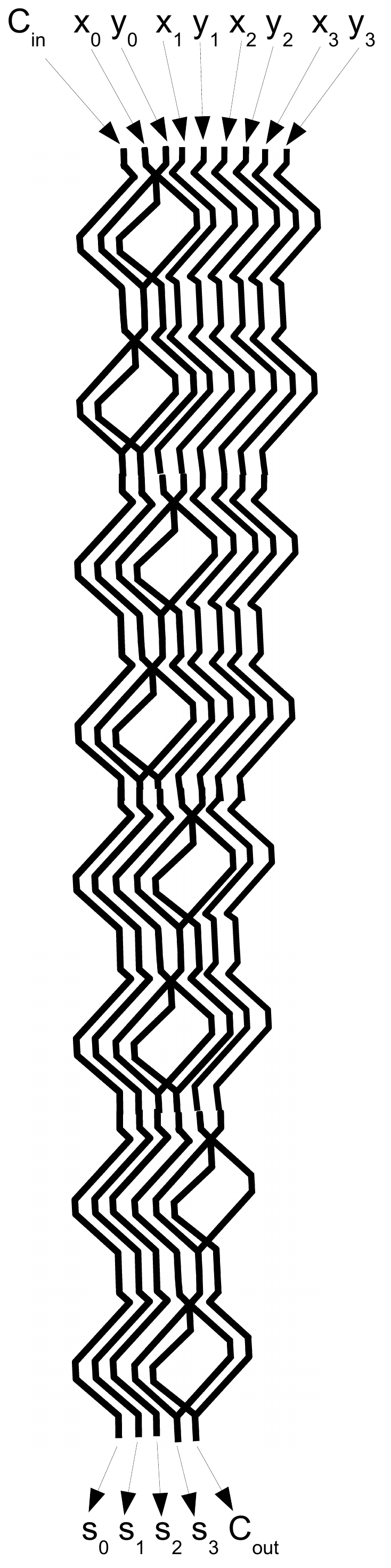}
\caption{Scheme of 4-bit full adder. Input bit-strings are $(x_0, x_1, x_2, x_3)$ and $(y_0, y_1, y_2, y_3)$ and their sum bit-string is $(s_0, s_1, s_2, s_3)$.}
\label{4bitadder}
\end{figure}

The full adder blocks  (Fig.~\ref{fulladderscheme}) can be straightforwardly cascaded into a many-bit adder.  An example of a 4-bit adder is shown in Fig.~\ref{4bitadder}. As we can see, an $n$-bit full adder can be made of $2n$ one-bit half-adders. Signals, represented by wave-fragments, are fully synchronised. Therefore, we can feed input data into the many-bit adder with period exceeding two wave-width, so excited heads of current signals do not interfere with refractory tails of previous signals. Assuming that a width of excitation wave-front in BZ system is circa 0.5-1~mm \cite{ginn2004microfluidic, davydov2002critical}, and the speed of propagation is about 1~mm per minute, we believe that many-bits full adder's clock frequency could be 0.01~Hz. With regards to a spatial complexity, to stay on the safe side we assume that one-bit full adder fits into $10 \times 3$~mm box, thus $n$-bit adder would take $10 \cdot n \times 3$~mm space. That is e.g. eight-bit adder implemented with fusion gates in sub-excitable BZ medium would be 8~cm by 0.3~cm in size.

\section{Discussion}
\label{discussion}

We proposed implementation of binary adders in architectures of homogeneous channels in  Belousov-Zhabotinsky (BZ) system.   A fusion gate is a key element of the architecture. The function of the fusion gate is to allow lonely excitation wave-fragments to pass along their original trajectories undisturbed but directing a new wave-fragment born after collision between two wave-fragments along the new trajectory. Localised wave-fragments exist only in sub-excitable BZ system.  Unfortunately, they are unstable: after some period of propagation the wave-fragments start expanding or collapsing. Thus to keep signals propagation without decay, we kept BZ medium in the excitable mode inside the channel. To prevent signals from exploding at the junction we kept the BZ medium in sub-excitable mode at the junctions between the channels.  If required, the  whole architecture can be kept in sub-excitable mode by enforcing periodic changes of excitability. 

In experimental laboratory conditions, the architecture can be implemented in two ways. We can make physical templates of the arithmetical circuits, where channels are depressions filled with BZ solution. Such approach would be robust, however the whole architecture would be non-reconfigurable. And we would still need to apply illumination to control excitability.   Using only light would be the most efficient and elegant approach.  If BZ medium is light-sensitive then generation and propagation of waves is controlled by illlumination. In the Oregonator model (\ref{equ:oregonator})  this control is expressed via constant $\phi$. This is a rate of inhibitor production proportional to intensity of illumination.  Thus the architecture of a computing circuit  can be projected onto the medium, in such a manner that channels are black and background is white. Then excitation waves will be confined to the channels because rest of the medium would be non-excitable.  Moreover, by dynamically varying illumination at the junctions we can implement polymorphic logical gates by controlling outcomes of inter-fragment collisions using the illumination level. For example, in~\cite{adamatzky2011polymorphic}  we demonstrated that how to switch the BZ collision gate between {\sc nor} to {\sc xnor} modes. 

\begin{figure}[!tbp]
\subfigure[]{\includegraphics[width=0.325\linewidth]{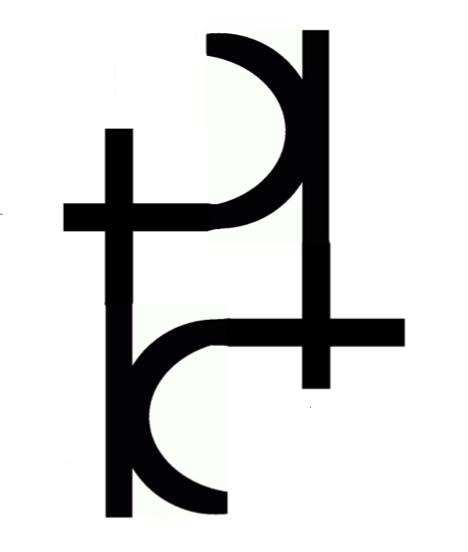}}
\subfigure[]{\includegraphics[width=0.325\linewidth]{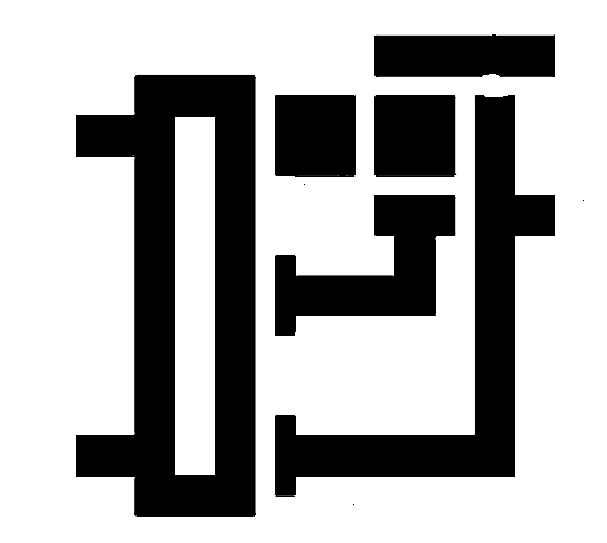}}
\subfigure[]{\includegraphics[width=0.325\linewidth]{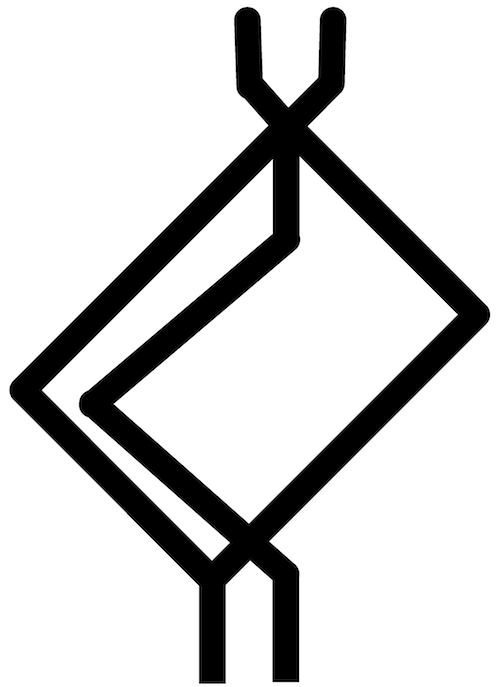}}
\caption{Architectures of three types of one-bit half-adders implemented in BZ system. (a)~Collision based design~\cite{adamatzky2010slime, costello2011towards}. (b)~Coincidence detectors and chemical diodes based design~\cite{suncrossover} and (c~) Wave-fragments fusion based design. } 
\label{comparison}
\end{figure}

Let us compare our design with previous implementations (Fig.~\ref{comparison}. Our previous design of  one-bit half-adder implemented in BZ system in computer model~\cite{adamatzky2010slime} and laboratory experiments~\cite{costello2011towards}  has four inputs (data signals are duplicated) and four junctions where wave-fragments interact (Fig.~\ref{comparison}a).  Half-adder implemented in computer models in~\cite{suncrossover} has two inputs and two outputs and seven junctions, or interactions loci (Fig.~\ref{comparison}b). The architecture~\cite{suncrossover} consists of six disconnected domains of `conductivity' and has seven junctions, which control propagation of excitation wave-fronts. The half-adder designed in our present paper (Fig.~\ref{comparison}c) has two input and two outputs, only three junctions between channels, and wave-fragment can interact only in two junctions. 

Obviously, one can find advantages in every implementation. For example, the first ever BZ half-adder design~\cite{adamatzky2010slime, costello2011towards} produces conjunction in addition to carry out and sum; and, the design~\cite{suncrossover} does not require a control of excitability at the junctions.  Yet our present architecture seems to be more optimal because it is made of a single domain of `conductivity' where excitation wave-fragments can interact only in two junctions. Also, speaking in terms of laboratory experiments each coincidence detector~\cite{gorecka2003t} and/or chemical diode~\cite{DBLP:journals/ijuc/IgarashiG11}, used in~\cite{suncrossover}, add to overall unreliability of the system because things can go wrong when excitation waves try to pass loci of discontinuity, because 
even a width of channel~\cite{DBLP:journals/ijuc/SzymanskiG10, kitahata2004slowing} strongly affects dynamics of wave-fronts.

There are three obvious directions of further studies: speed up, application domain, and extension to many-valued circuits. BZ circuits implemented in experimental laboratory conditions are slow, each cycle of computation takes quarter of an hour. To speed up the computation, we can implement the BZ adders in analog devices~\cite{asai2005analog},  molecular arrays~\cite{DBLP:journals/ijuc/Reif12} and at the nano-scale~\cite{DBLP:journals/ijuc/KonkoliW14}. With regards to the application domain, we envisage that our designs can be successfully integrated in embedded chemical sensing devices~\cite{DBLP:journals/ijuc/NakataIMI09}, \cite{DBLP:journals/ijuc/YoshikawaNIGGI09}. And yet another direction of future research could be in expanding our designs to many-valued logic circuits. In principle, we can implement three-valued gates with wave-fragments in BZ system~\cite{motoike2005three}, based on facilitation or inhibition between two subsequent excitation fronts, more work is required though to achieve cascading of these many-valued logic gates.

\bibliography{fusion_gate}

\end{document}